\documentstyle[preprint,prl,aps]{revtex}

\input epsf

\begin{document}
\title{Stable $^{85}\mbox{Rb}$ Bose-Einstein Condensates with Widely
  Tunable Interactions}
\author{S. L. Cornish, N. R. Claussen, J. L.
  Roberts, E. A. Cornell\cite{byline} and C. E. Wieman}
\address{JILA,
  National Institute of Standards and Technology and the University of
  Colorado, and the
  Department of Physics, University of Colorado, Boulder, Colorado
  80309-0440}
\date{\today}
\maketitle

\begin{abstract}
  Bose-Einstein condensation has been achieved in a magnetically
  trapped sample of $^{85}\mbox{Rb}$ atoms. Long-lived condensates of
  up to $10^4$ atoms have been produced by using a
  magnetic-field-induced Feshbach resonance to reverse the sign of the
  scattering length. This system provides many unique opportunities
  for the study of condensate physics. The variation of the scattering
  length near the resonance has been used to magnetically tune the
  condensate self-interaction energy over a very wide range. This
  range extended from very strong repulsive to large attractive
  self-interactions. When the interactions were switched from
  repulsive to attractive, the condensate shrank to below our
  resolution limit, and after $\sim\!5\,\mbox{ms}$ emitted a burst of
  high-energy atoms.
\end{abstract}


Atom-atom interactions have a profound influence on most of the
properties of Bose-Einstein condensation (BEC) in dilute alkali gases.
These interactions are well described in a mean-field model by a
self-interaction energy that depends only on the density of the
condensate $(n)$ and the s-wave scattering length $(a)$
\cite{Dalfovo1999a}.  Strong repulsive interactions produce stable
condensates with a size and shape determined by the self-interaction
energy. In contrast, attractive interactions ($a<0$) lead to a
condensate state where the number of atoms is limited to a small
critical value determined by the magnitude of $a$
\cite{Ruprecht1995a}. The scattering length also determines the
formation rate, the spectrum of collective excitations, the evolution
of the condensate phase, the coupling with the noncondensed atoms, and
other important properties.

In the vast majority of condensate experiments the scattering length
has been fixed at the outset by the choice of atom.  However, it was
proposed that the scattering length could be controlled by utilizing
the strong variation expected in the vicinity of a
magnetic-field-induced Feshbach resonance in collisions between cold
$(\sim\!\mu\mbox{K})$ alkali atoms \cite{Tiesinga1993a}. Recent
experiments on cold $^{85}\mbox{Rb}$ and Cs atoms and Na condensates
have demonstrated the variation of the scattering length via this
approach \cite{Roberts1998a,Courteille1998,Vuletic1999a,Inouye1998a}.
However, extraordinarily high inelastic losses in the Na condensates
were found to severely limit the extent to which the scattering length
could be varied and precluded an investigation of the interesting
negative scattering length regime \cite{Stenger1999c}.  These findings
prompted the subsequent proposal of several exotic coherent loss
processes that remain untested in other alkali species
\cite{Timmermans1999a,vanAbeelen1999a,Yurovsky1999a}.

Here we report the successful use of a Feshbach resonance to readily
vary the self-interaction of long-lived condensates over a large
range. In $^{85}\mbox{Rb}$ there exists a Feshbach resonance in
collisions between two atoms in the
$\mbox{F}\!=\!2,\,\mbox{m}_{\mbox{f}}\!=\!-2$ hyperfine ground state
at a magnetic field $B\!\sim\!155\,\mbox{Gauss(G)}$
\cite{Roberts1998a,Courteille1998a}.  Near this resonance the
scattering length varies dispersively as a function of magnetic field
and, in principle, can have any value between $-\,\infty$ and
$+\,\infty$ (see inset in Fig.~\ref{expplots}). This has allowed us to
reach novel regimes of condensate physics. These include producing
very large repulsive interactions $(n_{pk}|a|^{3}\!\simeq\!10^{-2})$,
where effects beyond the mean-field approximation should be readily
observable. We can also make transitions between repulsive and
attractive interactions (or vice-versa).  This now makes it possible
to study condensates in the negative scattering length regime,
including the anticipated ``collapse'' of the condensate
\cite{Kagan1997a}, with a level of control that has not been possible
in other experiments \cite{Bradley1997a}.  In fact, this ability to
change the sign of the scattering length is essential for the
existence of our $^{85}\mbox{Rb}$ condensate.  Away from the resonance
the large negative scattering length ($\simeq\!-400\,a_0$
\cite{Roberts1998a,Burke2000}) limits the maximum number of atoms in a
condensate to $\sim\!80$ \cite{Ruprecht1995a}. However, we have
produced condensates with up to $10^4$ atoms by operating in a region
of the Feshbach resonance where $a$ is positive.

The experimental apparatus was similar to the double magneto-optical
trap (MOT) system used in our earlier work \cite{Myatt1996a}. Atoms
collected in a vapor cell MOT were transferred to a second
MOT in a low-pressure chamber. Once sufficient atoms have accumulated
in the low-pressure MOT, both the MOT's were turned off and the atoms
were loaded into a purely magnetic, Ioffe-Pritchard ``baseball'' trap.
During the loading sequence the atom cloud was compressed, cooled and
optically pumped, resulting in a typical trapped sample of about
$3\times10^{8}$ $^{85}\mbox{Rb}$ atoms in the
$\mbox{F}\!=\!2,\,\mbox{m}_{\mbox{f}}\!=\!-2$ state at a temperature
of $T=45\,\mu\mbox{K}$.  The lifetime of atoms in the magnetic trap
due to collisions with background atoms was $\simeq450\,\mbox{s}$.

Forced radio-frequency evaporation was employed to cool the sample of
atoms. Unfortunately, $^{85}\mbox{Rb}$ is plagued with pitfalls for
the unwary evaporator familiar only with $^{87}\mbox{Rb}$ or Na. In
contrast to those atoms, the elastic collision cross section for
$^{85}\mbox{Rb}$ exhibits strong temperature dependence due to a zero
in the s-wave scattering cross section at a collision energy of
$E/k_{B}\simeq350\,\mu\mbox{K}$ \cite{Burke1998a}.  This decrease in
the elastic collision cross section with temperature means that the
standard practice of adiabatic compression to increase the initial
elastic collision rate does not work. $^{85}\mbox{Rb}$ also suffers
from unusually high inelastic collision rates. We recently
investigated these losses and observed a mixture of two and three-body
processes which varied with $B$ \cite{Roberts2000}. The overall
inelastic collision rate displayed several orders of magnitude
variation across the Feshbach resonance, with a dependence on $B$
similar to that of the elastic collision rate.  However, the inelastic
rate increased more rapidly than the elastic rate towards the peak of
the Feshbach resonance, and was found to be significantly lower in the
high field wing of the resonance than on the low field side. This
knowledge of the loss rates, together with the known field dependence
of the elastic cross section \cite{Roberts1998a,Burke2000}, has
enabled us to successfully devise an evaporation path to reach BEC.
This begins with evaporative cooling at a field $(B=250\,\mbox{G})$
well above the Feshbach resonance. To maintain a relatively low
density, and thereby minimize the inelastic losses, a relatively weak
trap is used. The low initial elastic collision rate means that about
$120\,\mbox{s}$ are needed to reach $T\simeq2\,\mu\mbox{K}$, putting a
stringent requirement on the trap lifetime. As the atoms cool, the
elastic rate increases and it becomes advantageous to trade some of
this increase for a reduced inelastic collision rate by moving to
$B=162.3\,\mbox{G}$ where the magnitude of the scattering length is
decreased. The remainder of the evaporation is performed at this field
(with a radial (axial) trap frequency of $17.5\,\mbox{Hz}$
$(6.8\,\mbox{Hz})$). In contrast to field values away from the
Feshbach resonance, the scattering length is positive at this field
and stable condensates may therefore be produced.

The density distribution of the trapped atom cloud was probed using
absorption imaging with a $10\,\mu\mbox{s}$ laser pulse
$1.6\,\mbox{ms}$ after the rapid $(\simeq0.2\,\mbox{ms})$ turn-off of
the magnetic trap. The shadow of the atom cloud was magnified by about
a factor of 10 and imaged onto a CCD array to determine the spatial
size and the number of atoms. The emergence of the BEC transition was
observed at $T\simeq15\,\mbox{nK}$.  Typically, we were able to
produce ``pure'' condensates of up to $10^4$ atoms with peak number
densities of $n_{pk}\simeq1\times10^{12}\mbox{cm}^{-3}$. The lifetime
of the condensate at $B=162.3\,\mbox{G}$ was about $10\,\mbox{s}$
\cite{sodium}. This lifetime is consistent with that expected from the
inelastic losses we have measured in cold thermal
clouds\cite{Roberts2000}.  It is notable that our evaporation
trajectory suffered a near-catastrophic decline prior to the
observation of the BEC transition. We approached the required BEC
phase space density at ~100 nK with about $10^{6}\,$atoms, but then
lost a factor of about $50$ in the atom number before the
characteristic two-component density distribution was visible. Over
this part of the trajectory, the cooling efficiency has become low
(and the phase space density remain approximately constant).  This is
because the mean free path is comparable to the cloud size and the
high density results in high losses. This situation improves when the
number of atoms becomes sufficiently low, and we are then able to
obtain a significant fraction of the atoms in a condensate.  The
number of atoms in the condensate after we reach this favorable
low-density regime is obviously a delicate balance between the elastic
(cooling) and inelastic (loss) collision processes in the cloud.  Both
of these are strongly field dependent near the Feshbach resonance.
Although we are able to decrease the loss rate by moving to higher
fields, the ratio of elastic to inelastic collisions actually
decreases and it becomes harder to form condensates. For example, at
$B=164.3\,\mbox{G}$ we can only produce condensates of a few thousand
atoms. Conversely, moving to lower fields does not help because we
reach the favorable low-density cooling regime at smaller numbers of
atoms.  This restriction together with the larger loss rate means that
at $B=160.3\,\mbox{G}$, for example, we are unable to form
condensates.

One of the features of the high inelastic loss rates reported in the
Na experiments was an anomalously high decay rate when the condensate
was swept rapidly through the Feshbach resonance \cite{Stenger1999c}.
In light of this work, it was essential to determine to what extent
the $^{85}\mbox{Rb}$ condensate was perturbed in being swept across
the Feshbach peak during the trap turn-off.  These measurements also
provide an additional test of coherent loss mechanisms such as those
in references \cite{Timmermans1999a,vanAbeelen1999a,Yurovsky1999a}.
We applied a linear ramp to the current in the baseball coil to sweep
the magnetic field experienced by the atoms from $B=162.3\,\mbox{G}$
across the Feshbach peak to $B\simeq132\,\mbox{G}$ and then
immediately turned off the trap and imaged the atom cloud. From the
images we determined the fraction of condensate atoms lost as a
function of the inverse ramp speed (Fig.~\ref{losses}). The loss for
the fastest ramp, which corresponds to the direct turn-off of the
magnetic trap, is less than $9\,\%$. This was determined in a separate
experiment where the condensate was imaged directly in the magnetic
trap both before and after the ramp. For comparison, the experiment
was repeated using a cloud of thermal atoms much hotter than the BEC
transition temperature. The results for the thermal atoms are
consistent with the known inelastic loss rates in the vicinity of the
Feshbach resonance \cite{Roberts2000}. The strong and poorly
characterized temperature dependence of these known loss rates near
the Feshbach peak makes it difficult to determine what fraction of the
observed condensate loss can be attributed to the usual inelastic loss
processes and we cannot, therefore, rule out a coherent aspect to the
loss process.  There have been several models of coherent loss
processes put forward to explain the corresponding sodium
results\cite{Timmermans1999a,vanAbeelen1999a,Yurovsky1999a}.  However,
these calculations are based on the Timmermans theory
\cite{Timmermans1999a} of coupled atomic and molecular
Gross-Pitaevskii equations which is unlikely to be applicable to the
conditions of the present experiment \cite{Eric}.

We also changed the self-interaction energy by varying the magnetic
field and observed the resulting change in the condensate shape. By
applying a linear ramp to the magnetic field we have varied the
magnitude of $a$ in the condensate by almost three orders of
magnitude. The duration of this ramp was sufficiently long
$(500\,\mbox{ms})$ to ensure that the condensate responded
adiabatically. Fig.~\ref{exppics} shows a series of condensate images
for various magnetic fields.  They illustrate how we are able to
easily change $a$ over a very wide range of positive values. Moving
towards the Feshbach peak the condensate size increases due to the
increased self-interaction energy. The density distribution approaches
the parabolic distribution with an aspect ratio of
$\lambda_{TF}\!=\!\omega_{z}/\omega_{r}$ expected in the TF regime
\cite{Baym1996a}. Moving in the opposite direction the cloud size
becomes smaller than our 7 micron resolution limit shortly before we
reach the noninteracting limit where the condensate density
distribution is a Gaussian whose dimensions are set by the harmonic
oscillator lengths ($l_{i}=(\hbar/m\omega_{i})^{1/2}$ where $i=r,z$)
\cite{Baym1996a}.  We took condensate images similar to those shown at
many field values between 156 and 166 G.  From the images the full
widths at half-maximum (FWHM) of the column density distributions were
determined.  The scattering length was then derived by assuming a TF
column density distribution with the same FWHM
$(\propto\!(Na)^{1/5})$. In Fig.~\ref{expplots} we plot the scattering
length derived in this manner versus the magnetic field.  It shows
that these values agree with the predicted field dependence of the
Feshbach resonance.

The ability to tune the atom-atom interactions in a condensate
presents several exciting avenues for future research. One is to
explore the breakdown of the dilute-gas approximation near the
Feshbach peak. The lifetime of the condensate decreases with larger a,
but for a lifetime of about 100 ms, which is sufficient for many
experiments, we have created static condensates with
$n_{pk}|a|^{3}\simeq10^{-2}$. For such values, effects beyond the mean
field approximation, such as shifts in the frequencies of the
collective excitations \cite{Pitaevskii1998a}, are about 10\%.

A second avenue is the behavior of the condensate when the scattering
length becomes negative.  When we increased the magnetic field beyond
$B=166.8\,\mbox{G}$, where $a$ was expected to change sign, a sudden
departure from the smooth behavior in Figs.~\ref{expplots}
and~\ref{exppics} was observed.  As $a$ was decreased the condensate
width decreased and then about 5 ms seconds after the change in sign
there was a sudden explosion that ejected a large fraction of the
condensate.  This left a small observable remnant condensate
surrounded by a ``hot'' cloud at a temperature on the order of 100 nK.
These preliminary results on switching from repulsive to attractive
interactions suggest a violent, but highly reproducible, destruction
of the condensate. The ability to control the precise moment of the
onset of $a<0$ instability is a distinct advantage over existing
methods for studying this regime which rely on analyzing ensemble
averages of ``post-collapse'' condensates \cite{Bradley1997a}. The
dynamical response of the condensate to a sudden change in the sign of
the interactions can now be investigated in a controlled manner,
probing the rich physics of this dramatic condensate ``collapse''
process.

We are pleased to acknowledge useful discussions with Murray Holland,
Jim Burke, Josh Milstein and Marco Prevedelli. This research has been
supported by the NSF and ONR. One of us (S. L.  Cornish) acknowledges
the support of a Lindemann Fellowship.


\begin{figure}
\caption{Scattering length in units of the Bohr radius $(a_0)$ as a
  function of the magnetic field. The data are derived from the
  condensate widths. The solid line illustrates the expected shape of
  the Feshbach resonance using a peak position and resonance width
  consistent with our previous measurements
  {\protect\cite{Roberts1998a,Roberts2000}}. For reference, the shape
  of the full resonance has been included in the inset.}
\label{expplots}
\end{figure}

\begin{figure}
\caption{Fraction of atoms lost following a rapid sweep of the
  magnetic field through the peak of the Feshbach resonance as a
  function of the inverse speed of the field ramp. Data are shown for
  a condensate $(\bullet)$ with a peak density of
  $n_{pk}=1.0\times10^{12}\,\mbox{cm}^{-3}$ and for a thermal cloud
  $(\circ)$ with a temperature $T=430\,\mbox{nK}$ and a peak density
  of $n_{pk}=4.5\times10^{11}\,\mbox{cm}^{-3}$.}
\label{losses}
\end{figure}

\begin{figure}
\caption{(color) False color images and horizontal cross sections of
  the condensate column density for various magnetic fields.  The
  condensate number was varied to maintain an optical depth (OD) of
  $\sim1.5$.  The magnetic field values are (a) $B=165.2\,\mbox{G}$, (b)
  $B=162.3\,\mbox{G}$,(c) $B=158.4\,\mbox{G}$,
  (d) $B=157.2\,\mbox{G}$, (e)
  $B=156.4\,\mbox{G}$.}
\label{exppics}
\end{figure}

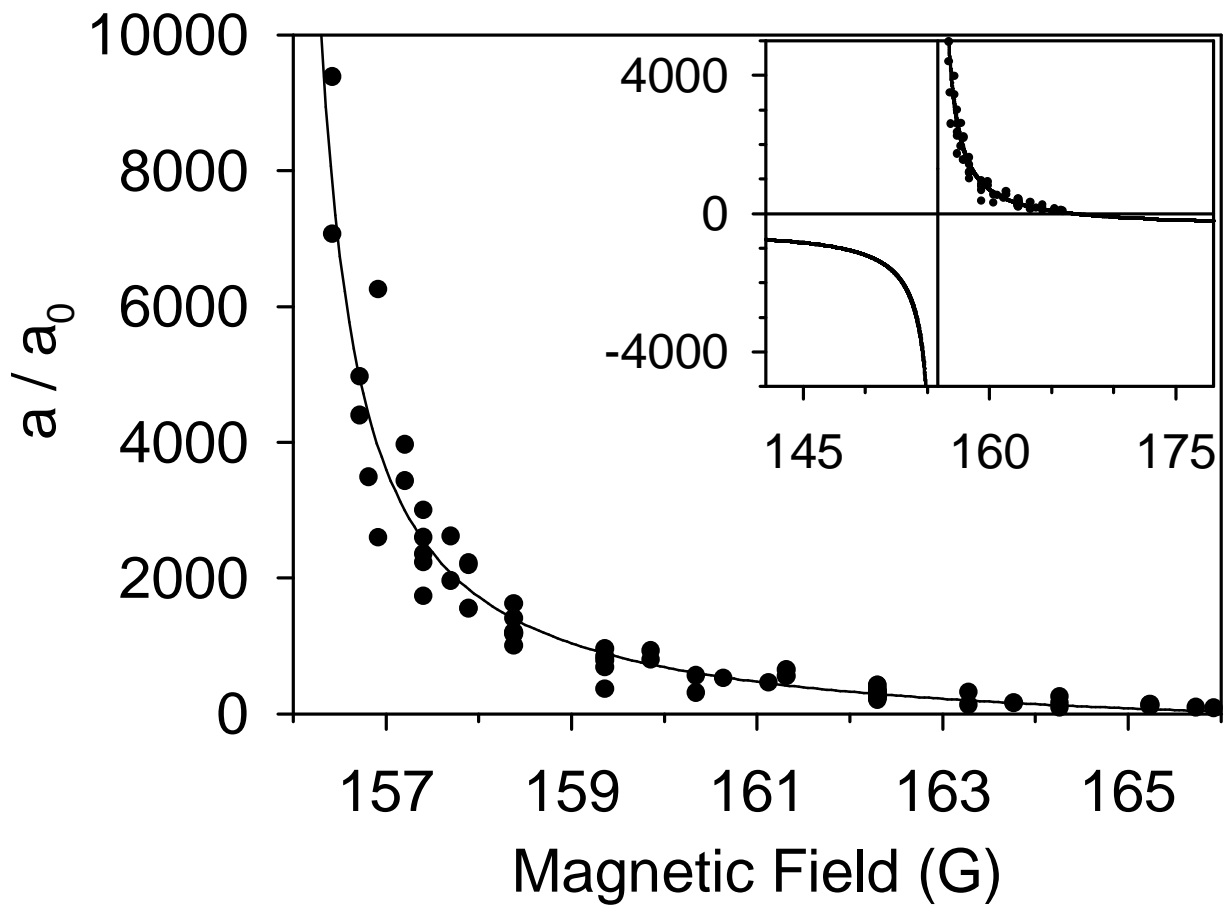
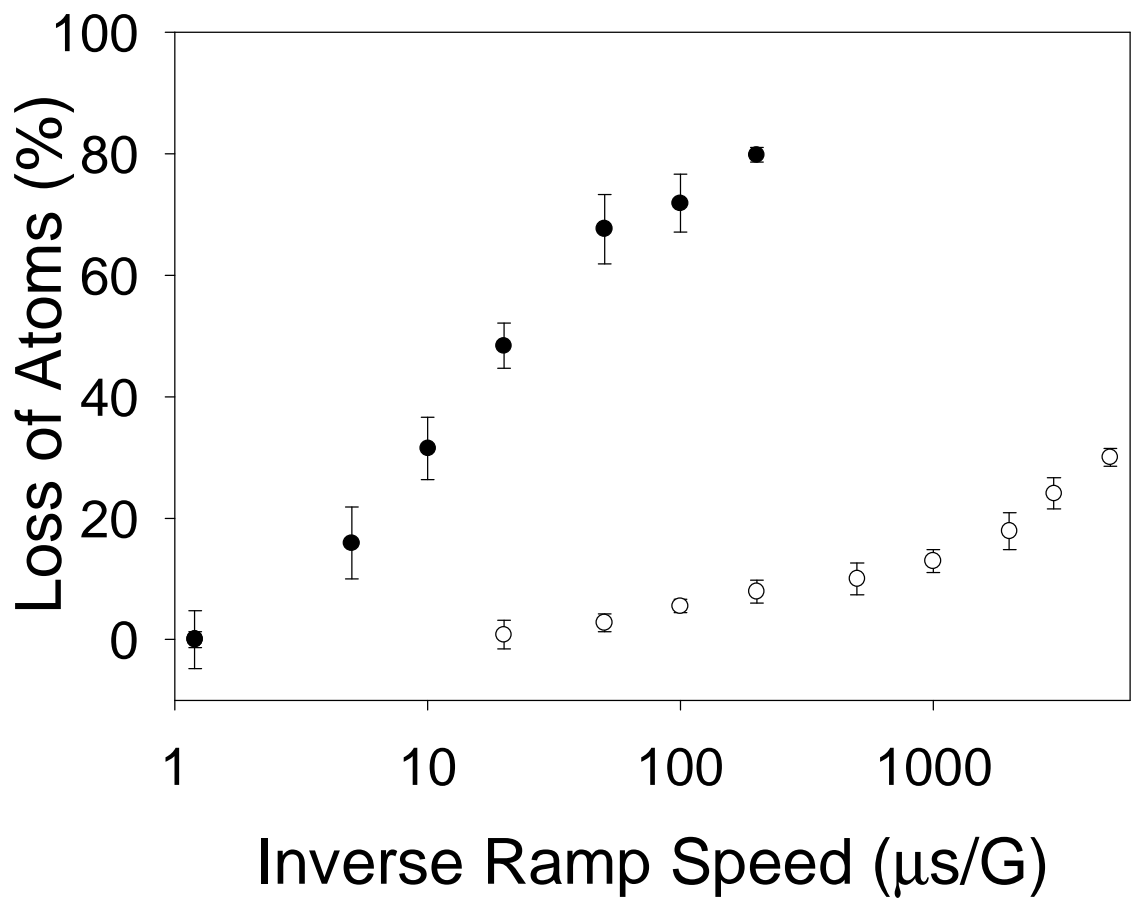
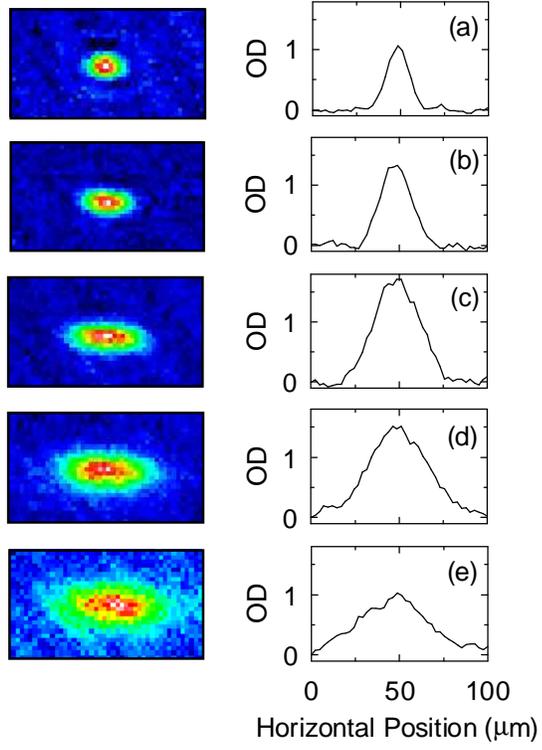
\end{document}